\documentclass[aps,prd,twocolumn,nofootinbib,superscriptaddress,amssymb]{revtex4}
\usepackage{color}
\usepackage{epsfig}
\usepackage{graphicx}
\usepackage{epstopdf}
\usepackage{mathtools}
\usepackage{bbold}
\def\be{\begin{equation}}
\def\ee{\end{equation}}

\begin{document}

\title{Passing through the Firewall}

\author{Erik Verlinde}

\affiliation{Institute for Theoretical Physics, University of Amsterdam, Amsterdam, The Netherlands}

\def\spc{\hspace{.5pt}}

\author{Herman Verlinde}
\affiliation{Department of Physics, Princeton University, Princeton, NJ 08544, USA}

\date{\today}

\begin{abstract}

We propose that black hole information is encoded in non-local correlations
between microscopic interior and exterior degrees of freedom. 
We give a simple qubit representation of this proposal, and show herein that for every black hole state, the apparent firewall can be removed 
via a universal, state independent unitary transformation ${\bf U}_{{\! }_{\rm QT}}$.  
A central element in our discussion is the distinction between~virtual qubits, which are in a specified vacuum state, and real qubits, that carry the free quantum~information of the black hole. We outline how our proposal may be realized in AdS/CFT.

\end{abstract}

\def\be{\begin{equation}}
\def\ee{\end{equation}}
\maketitle
\def\mathbi#1{\textbf{\em #1}} 
\def\la{\langle}
\def\bea{\begin{eqnarray}}
\def\eea{\end{eqnarray}}
\def\is{\! & \! = \! & \!}
\def\half{{\textstyle{\frac 12}}}

\def\ba{\begin{eqnarray}}
\def\ea{\end{eqnarray}}





\def\be{\bea}
\def\ee{\eea}
\newcommand{\smpc}{\hspace{.5pt}}
\def\ra{\bigr\rangle}
\def\la{\bigl\langle}
\def\nspc{\!\spc\smpc}
\def\tr{{\rm tr}}
\def\bh{{\mbox{\fontsize{7pt}{.7pt}{$BH$}}}}

\def\Aa{{\mbox{\scriptsize \smpc \sc a}}}
\def\bfC{\mbox{{\textbf C}}}
\def\bC{\alpha} 
\def\nonu{\nonumber}
\def\sC{{\mbox{\scriptsize {\smpc \sc sc}}}}
\addtolength{\baselineskip}{.3mm}
\addtolength{\parskip}{.3mm}
\renewcommand\Large{\fontsize{15.5}{16}\selectfont}
\def\ra{\bigr\rangle}
\def\la{\bigl\langle}
\def\li{\bigl|\spc}
\def\ri{\bigr |\spc}

\def\hf{\textstyle \frac 1 2}

\def\cE{{\mbox{\tiny \nspc $\callE$}}}
\def\Ee{{\mbox{\scriptsize \smpc \sc e}}}
\def\Bb{{\raisebox{-.2pt}{\scriptsize \smpc \sc b}}}
\def\Zz{{\raisebox{-.2pt}{\scriptsize \smpc \sc z}}}

\def\Bbt{{\raisebox{-.2pt}{\scriptsize \smpc $\tilde{\mbox{\sc b}}$}}}
\def\Hh{{\mbox{\scriptsize \smpc \sc h}}}
\def\Aa{{\mbox{\scriptsize \smpc \sc a}}}

\def\BH{{\mbox{\scriptsize \smpc \sc bh}}}

\setcounter{tocdepth}{2}

\subsection{Introduction} 
\vspace{-3mm}

The Bekenstein-Hawking entropy counts the quantum states associated to a  black hole horizon \cite{bh}. This famous formula seems to
suggest that all black hole quantum information is hidden behind the event horizon and that all of it has to escape during the evaporation process \cite{hologram}. 
A second, alternative interpretation of the B\spc-H entropy is that it merely represents the virtual entanglement entropy of the unique smooth vacuum
state across the horizon. This local vacuum state does not store any evident quantum information,  neither do the near-horizon
Hawking particles. Both interpretations of the B-H formula have their merits and appeal, but there is evident tension between the two. Moreover, both lead to conceptual problems,  each of a different kind: the second interpretation leads to apparent information loss \cite{infoloss,mathur}, the first interpretation leads to an apparent firewall \cite{amps, followup}. 

In this paper we introduce a third interpretation of the B-H entropy formula,  which incorporates a balanced combination of both ideas.  
In our proposal, which we will call balanced holography, the microscopic quantum information is shared evenly between the interior and exterior of the black hole.
Since the two sides combined contain twice the B\spc-H entropy, this leaves extra room to arrange the quantum degrees of freedom in way that avoids  inconsistent or paradoxical conclusions.

An important distinction will be made between two kinds of quantum information. The entanglement contained in the local vacuum is of a different nature than that between two randomly prepared and spatially separated real quantum systems. The qubits responsible for vacuum entanglement are, when left unperturbed by local probes, in a unique ground state. Hence, they are `virtual' or `confined'. They are distinct from the `real' or `logical' qubits that carry freely adjustable quantum information about the micro-state, and that are determined by the initial state from which the black hole was formed. The difference between real and virtual qubits thus lies in the amount of free quantum information that they contain about the micro states corresponding to a given macroscopic black hole.
The purpose of this note is to show that the crux of the firewall argument can be neutralized, once the distinction between these two types of quantum information is taken into account.

\vspace{-3mm}

\subsection{Physical Principles} 
\vspace{-3mm}

In this work we adopt the basic physical principles that {i)}~quantum mechanics holds, 
ii) a black hole has microscopic entropy  $S_{{}_{\! \rm B\nspc H}}\! =\! \frac 1 4 A$, and iii) there are no firewalls,
iv) no observer sees any macroscopic violation of locality. 

We supplement these principles by two additional ground rules. 
For clarity of presentation, we will ignore peripheral subtleties, and
express our physical assumptions  and reasoning using the crisp 
but slightly idealized language of qubits \cite{hp, bitmodels}. This simplification is helpful but presumably not essential.

A small word of caution: a qubit can be coded in many physical forms. It may be carried by a localized mode, by hidden microscopic degree of freedom \cite{fuzzball}, or by non-local phase correlations between two or more distant modes.
We will try to be clear which physical realization we are assigning to each element.

\vspace{.5mm}
\begin{center}
{\it Rule I: Maximal entanglement}
\end{center}
\vspace{-2mm}

Consider a black hole with entropy equal to $N$ qubits. Its internal Hilbert space is $2^{N}$ dimensional. 
Suppose the black hole interior region H is maximally entangled with its environment E.  
The total quantum entanglement is then quantified via 
\be
\label{ents}
\mbox{entanglement entropy} = S_{{}_{\nspc \rm B\nspc H}} = N \log 2 
\ee 

There are two types of contributions to (\ref{ents}). First, there is local entanglement due to virtual Hawking pairs.
The associated `virtual qubits' are in unique EPR pairs (otherwise they can not appear from the vacuum)  and do not carry any quantum information about the black hole micro state.
This virtual entanglement is proportional to the area, with a normalization constant that depends on the short distance cut off \cite{sred, braunstein}. 
 Secondly, there is the actual black hole information, which 
is contained in `real' or `logical qubits'. These logical qubits  carry medium and long range entanglement, that over time can be decoded from the on-shell asymptotic 
Hawking radiation \cite{page}. An important part of the challenge is to reconcile the necessary presence of both types of
entanglement with the fact that their sum can never exceed the value (\ref{ents}):
the two types of entanglement are in competition and one of the key questions is how the balance between them depends on the age of the black hole relative to the Page time.

\def\Rr{{\mbox{\scriptsize \smpc \sc r}}}
Our first hypothesis is that situation (\ref{ents}) is typical:

\smallskip
\medskip

\parbox{8.2cm}{
\addtolength{\baselineskip}{.7mm}
 \it A typical quantum black hole, soon after it is formed,  is close to maximally entangled with its environment.  }
\medskip
\smallskip

\noindent
Various pieces of evidence support this postulate. The most compelling clue is provided by the Ryu-Takayanagi entropy formula \cite{RT},
which identifies the microscopic entanglement entropy, defined via the CFT, with $1/4$ of the area of the holographic entanglement surface in AdS.
When applied in presence of an AdS black hole, it indicates that its Bekenstein-Hawking entropy also quantifies the entanglement across the horizon.

As  emphasized  in \cite{van},
saturation of the B\spc -H entropy bound across an entanglement surface is the likely source of continuity of space.  A black hole interior described by 
a pure state is a singular notion: it would have a firewall. Instead, the black hole interior is a fast scrambler that promptly thermalizes:
the transition from gravitational collapse towards  a maximally mixed interior state takes of order the scrambling time $t_S \simeq M \log M$.

\begin{center}
{\it Rule II:  Balanced holography}
\end{center}
\vspace{-1mm}

We define the {\it entangled environment} E of a young black hole as the $2^N$ dimensional Hilbert space spanned by all states that are
entangled with the black hole interior  H.   Hence for now, E is a purely information theoretic construct. 
A general maximally entangled state of the  interior H and  exterior E takes the form
\be
\label{genstate}
\li \Psi \ra = \sum_{i, j}  \alpha_{ij}\, \li \spc i \spc \ra_{\! \Hh} \li \spc j \spc \ra_{\! \Ee}
\ee
where $i,j$ run from 1 to $2^N$, and $\alpha_{ij}$ are a priori arbitrary complex amplitudes.
The total Hilbert space ${\cal H}_{\Hh} \otimes {\cal H}_{\Ee}$ is $2^{2N}$ dimensional: the 
black hole engages $2N$ qubits. 
If all states  (\ref{genstate}) were equally accessible,
 the total thermodynamic entropy of the combined system H and E would  overshoot the B\spc-H bound by a factor of 2.

There is no a a priori physical principle, however, that prescribes that every state of the form (\ref{genstate}) 
must be reachable via  time-evolution from some reasonable initial state. The {\it interior} black hole dynamics is maximally ergodic, but the combined system 
H and E is not: when unperturbed by outside influences, it occupies only 
a small subset of its Hilbert space. Physical properties such as the total energy and the local energy distribution of the state 
depend on the coefficients $\alpha_{ij}$.
The quantum states associated to undisturbed (young) black hole space times satisfy local conditions that distinguish
vacuum states from excited states. The number of such black hole vacuum states is counted by the B\spc-H formula.

So our second postulate is that among all states of the form~(\ref{genstate}),
only a small subset are  physical black hole vacuum states

\smallskip
\medskip

\hspace{-1mm}\parbox{8.2cm}{\addtolength{\baselineskip}{.7mm}
\it The physical Hilbert space of a young black hole and its entangled environment E is $e^{S_\BH} = 2^N$ dimensional.}
\medskip
\smallskip

We call this principle `balanced holography', because when combined with the first postulate,  it implies that black hole information is not stored
inside the horizon, but carried by 
correlations between an equal number of internal and external qubits. For young black holes, we imagine that these `logical' qubits are shared 
by hidden interior and exterior microscopic degrees of freedom (c.f. \cite{fuzzball}), extending out into the zone region.

The real black hole information thus occupies only half the total number of qubits of H and E.
The remaining $N$ qubits represent virtual quantum fluctuations, and are required to be in a specific vacuum state, whose precise form is determined by the microscopic theory.
Excitations of the virtual qubits are allowed states in the total Hilbert space, but their local and total energy distribution is macroscopically distinct from the
typical state in the thermal ensemble of black hole states with a given mass and horizon area. 

\smallskip
 
These are our two additional postulates. They are complementary: neither would be consistent without the other. Note that a corollary of rule I is that there is no fundamental distinction between the interior state 
of a young or old black hole: both are maximally mixed, and should be able to support 
a smooth interior geometry. 

\vspace{-3mm}

\subsection{Vacuum conditions. }

\vspace{-3mm}

Balanced holography posits that among all $2^{2N}$  independent states in  ${\cal H}_\Hh \otimes {\cal H}_\Ee$, only $2^N$ are physical black hole states.
This means that the complete quantum state of the black hole and its environment takes the form 
\be
\label{gnstate}
\li \Psi \ra = \sum_{i}  \alpha_{i}\, \li \spc i \spc \ra_{\! \Hh} \li \spc i \spc \ra_{\! \Ee}\, .
\ee
where we used the freedom to relabel our interior black hole states so that they carry the same label $i$ as the environment states. The black hole quantum information is thus stored non-locally in the $2^N$ independent complex amplitudes $\alpha_i$, and is evenly divided over the interior and the exterior region of the horizon.  

Because the physical states $\li \Psi\ra \in {\cal H}_\Hh \otimes {\cal H}_\Ee$ of the form (\ref{gnstate}) only carry $N$ independent qubits of quantum information, they must satisfy $N$  conditions, each removing one of the original $2N$ qubits. The form of  these conditions is determined by the microscopic theory.  The $2N$ qubits are thus evenly divided into two types: there are $N$ real or `logical' qubits, which are free to take any value, 
and $N$ virtual qubits, which are in their vacuum state. 

Maximal entanglement between H and E implies that both regions have an equal number of real and virtual qubits.
Moreover, a typical qubit of each type is necessarily stored in correlations across the horizon. To see this,
consider one internal qubit. It is maximally entangled with an external qubit. Statistically, the  qubit pair contains one logical
 and one virtual qubit, and  each is evenly divided over the interior and exterior.
This balance holds within a statistical variance of order $1/{\sqrt{2^N}}$. 

Via this general reasoning, we deduce that a typical entangled qubit pair in a balanced black hole state takes 
the general form  (c.f. eq (\ref{gnstate}) )
\be \label{xtops}
 \li \Psi\ra =  \alpha_0\spc 
\li 0 \ra_{\! \Hh} \li 0 \ra_{\!\Ee} + \alpha_1\spc 
\nspc \li 1 \ra_{\! \Hh} \li 1 \ra_{\! \Ee},
\ee
where  $\alpha_{0}$  and $\alpha_1$ are arbitrary complex amplitudes. 
This state is maximally entangled if $|\alpha_{0}|^2  = |\alpha_1|^2 = 1/2$.
The entangled qubit pair of the form (\ref{xtops}) spans a two-dimensional subspace of all possible two qubit states, and thus carries one qubit of information. This logical qubit is stored in a non-local way and is shared between the horizon and the environment. The pair (\ref{xtops}) also carries one virtual qubit, which can be thought of carrying the information of
the spatial wavefunction of the logical qubit. If the virtual qubit is in its ground state, the internal and external qubit values are identical as in (\ref{xtops}), 
if it is in an excited state, they are opposite.

Let us make two simple but important remarks. The first comment is 
that the entanglement between H and E, and the bi-local nature of the virtual and logical qubits, can be removed by a single elementary logical operation: the controlled-not or CNOT, defined by flipping the first qubit provided the second qubit reads out as 1. Applying this CNOT operation to (\ref{xtops}) gives 
\be
\label{CNOT}
{\bf U}_{{\! }_{\rm CNOT}} \li \Psi\ra
= \li 0\ra_{\! \Hh}  \bigl( \alpha_0 \li 0 \ra _{\! \Ee} + \alpha_1\li1 \ra_{\! \Ee}\bigr)
\ee
The virtual and logical qubit are now manifest: the virtual qubit is mapped into the horizon state, while the environment qubit has become identified with the logical qubit. After the CNOT operation the environment thus contains all quantum information. 

Physically, the CNOT operation executes a unitary quantum teleportation protocol: it measures the exterior qubit, and then depending the outcome  it performs a 
rotation on a interior qubit. Thanks to the long range entanglement, this procedure transports coherent
quantum information from the interior to the exterior. The teleportation protocol is unitary, 
because there is no projection involved on a given  outcome of the measurement.

Our second remark is that each qubit can be seen as a represention of a pair of Dirac oscillators. We will denote them by $\mathbf{h}$ and $\mathbf{h}^\dagger$ for the horizon qubit and by $\mathbf{e}$ and $\mathbf{e}^\dagger$ for the environment qubit. The Dirac oscillators obey 
\be
\{\mathbf{h},\mathbf{h}^\dagger\}=1,\qquad \mbox{and}\qquad \mathbf{h}^2= \mathbf{h}^{\dagger\,2}=0,
\ee 
and similarly for $\mathbf{e}$ and $\mathbf{e}^\dagger$. 
The state $\li 0 \ra_{\! \Hh} $ represents the ground state of $\mathbf{h}$, and $\li 1 \ra_{\! \Hh}$ its excited state
\be
\mathbf{h} \li 0 \ra_{\! \Hh} = 0, \qquad \mathbf{h}^\dagger \li 0 \ra_{\! \Hh} = \li 1 \ra_{\! \Hh}.
\ee 

Let us now combine the above two comments.
We would like to be able to write all balanced two qubit states (\ref{xtops}) in the form of a vacuum state.
This means they all should obey similar vacuum conditions as the state $\li 0 \ra_{\! \Hh}$. 
It is straightforward to find these conditions by making use of the
CNOT operation (\ref{CNOT}). Note that ${\bf U}_{{\! }_{\rm CNOT}}$ may be expressed with the help of the Dirac oscillators as 
\be
{\bf U}_{{\! }_{\rm CNOT}} = (\mathbf{h}+\mathbf{h}^\dagger)^{n_e},\qquad \mbox{where} \qquad n_e =  \mathbf{e}^\dagger \mathbf{e}  
\ee
is the Dirac fermion number for the external qubit. 

The states $\li \Psi\ra$ given in (\ref{xtops}) are ground states for the Dirac operators $\mathbf{a}$ and $\mathbf{a}^\dagger$ associated to the virtual qubits. These oscillators $\mathbf{a}$ and $\mathbf{a}^\dagger$ are obtained 
by conjugating $\mathbf{h}$ and $\mathbf{h}^\dagger$ with ${\bf U}_{{\! }_{\rm CNOT}}$. We have 
\be
\label{acnot}
\mathbf{a}={\bf U}^{-1}_{{\! }_{\rm CNOT}}\mathbf{h} {\bf U}_{{\! }_{\rm CNOT}}.
\ee
Explicitly, one finds $\mathbf{a} =  \mathbf{h}(1-\mathbf{e}^\dagger \mathbf{e})+\mathbf{h}^\dagger\mathbf{e}^\dagger \mathbf{e}$. The fact that the virtual qubit carried by the state (\ref{xtops}) is in its ground state translates into the vacuum condition
\be
\label{vaccond}
\mathbf{a}  \li \Psi\ra = 0.
\ee
So we have succeeded in identifying all physical states (\ref{xtops}) with a local vacuum state. Moreover, the construction of the annihilation mode $\mathbf{a}$ is 
state independent: it works for all states of the form (\ref{xtops}). This observation is the key to the resolution of the firewall paradox.

\vspace{-2mm}

\subsection{Removing the Firewall}

\vspace{-2mm}

The original firewall argument, in short, goes as follows. Consider a quantum black hole for which  every interior qubit is entangled with a radiation qubit. 
Now, one qubit leaves the horizon through the Hawking process.  In order for an infalling observer to experience no drama, the qubit must form a unique EPR pair with another interior
qubit. This EPR pairing, however, is forbidden by monogamy of entanglement.


Let us revisit the argument in the context of balanced holography.
Consider  the following quartet of qubits
\be
\li s_1,\nspc s_2 \ra_{\!{}_\Hh}  \is \mbox{two interior qubits ${\bf h}_1$ and ${\bf h}_2$ inside H}\nonumber \\[-1mm]
\li s_3,\nspc s_4 \ra_{\!{}_\Ee} \is \mbox{two external qubits ${\bf e}_3$ and ${\bf e}_4$ inside E}\nonumber
\ee
Here $s_i$  take the values 0 or 1. We assume that the four qubits are in a maximally entangled state.
The Hilbert space of the quartet (${\bf h_1}, {\bf h_2}, {\bf e}_3, {\bf e}_4$) is $16$ dimensional,
and thus a priori, there are 16 independent basis states.
However,  we need to impose the balanced holography constraint, which reduces the number of allowed basis states to four.
Physical states can all be put in the form
\be
\label{newamps}
\li \Psi\ra=\sum_{s,r} \alpha_{sr}\, \li s,r\ra_{\! \Hh}\,  \li s,r\ra_{\! \Ee}
\ee
where $s,r=0,1$ and $\alpha_{r,s}$ are arbitrary complex amplitudes.  
The state (\ref{newamps}) is maximally entangled if $|\alpha_{s,r}|^2 = 1/4$  for all $s,r$. 

The Hilbert space of balanced four qubit states of the form (\ref{newamps}) factorizes into the tensor product of two 2-qubit Hilbert spaces of the balanced form (\ref{xtops}).
This fact will have some immediate practical use below.

The firewall argument still seems to apply to (\ref{newamps}): both interior qubits have their EPR partners outside of H.
So it looks like no qubit can leave the black hole while being in a unique EPR pair with another qubit inside of H. Namely, a smooth horizon state is expected to look like
\be
\label{nvac}
\li \Psi_0\ra=\sum_{s,r} \alpha_{sr}\, \li \mathbf{vac}\ra_{\! \Hh}\,  \li s,r\ra_{\! \Ee}
\ee
where $|\spc \mathbf{vac} \spc \rangle_{\! \Hh}$ denotes the vacuum  state
\be
\label{vac}
\li\spc \mathbf{vac}\spc \ra_{\! \Hh} 
\is {1\over \sqrt{1+v^2}}
\sum_{s} \, v^s\, \li s , s\ra_{\! \Hh}.
\ee
Here we included a possible Boltzmann factor $v$.
The states in eqs (\ref{newamps}) and (\ref{nvac}) look very different, so $|\Psi\rangle$ seems to describe a black hole with a singular horizon. 

A natural way to evade this unphysical conclusion is to instate the complementarity principle \cite{susskindcompl} that
$|\Psi\rangle$ and $|\Psi_0\rangle$ in fact do describe the same state, but seen by different observers:
the infalling observer experiences state (\ref{nvac}) and the outside observer sees state (\ref{newamps}).
The only way in which this could work, however, is if there exists a universal unitary operator  ${\mathbf U}_{{\! }_{\rm QT}}$
such that
\be
\label{nqtop}
\li \Psi_0 \ra \is \spc  {\mathbf U}_{{\! }_{\rm QT}} \spc \li \Psi\spc \ra \, .
\ee
To evade the monogamy obstruction, the operator ${\mathbf U}_{{\! }_{\rm QT}}$ must execute  unitary quantum teleportation protocol, or entanglement swap,  that transports 
all free quantum information from the interior H to the exterior E, and puts the internal qubits ${\bf h}_1$ and ${\bf h}_2$ into a unique state. 

Entanglement between two subsystems depends on the particular quantum state of the total system. So at first it may seem impossible to find a unitary operator ${\mathbf U}_{{\! }_{\rm QT}}$ that obeys (\ref{nqtop}) without introducing unacceptable non-linearity or state dependence. Via our discussion in subsection C, however, it is easy to see
that balanced holography in fact does allow for the existence of a universal swap operator ${\mathbf U}_{{\! }_{\rm QT}}$  with the property (\ref{nqtop}).
This basically follows from a counting argument: the Hilbert space of balanced black hole states (\ref{newamps}) has the same dimension as the Hilbert space
of vacuum states of the form (\ref{nvac}). So the two Hilbert spaces are unitarily equivalent. 
Indeed, as we will see, the 4-qubit operator ${\mathbf U}_{{\! }_{\rm QT}}$ is given by the tensor product of two 2-qubit CNOT operations ${\mathbf U}_{{\! }_{\rm CNOT}}$, described in subsection C, combined with a Bogoljubov transformation. 
 
The firewall can now be removed, because according to this complementarity principle the local operators by which the infalling observer tests whether  (\ref{newamps}) are vacuum states or not, are not given ${\bf h_1}, {\bf h_2}$  but by their image under conjugation with the unitary swap operation ${\mathbf U}_{{\! }_{\rm QT}}$ 
\be
{\mathbf b_i} =  {\bf U}^{-1}_{{\! }_{\rm QT}}\mathbf{h_i} {\bf U}_{{\! }_{\rm QT}}.
\ee
The usual tests, like measuring the entanglement of the virtual qubits across the horizon, will lead her to conclude that indeed all states of the form $(\ref{newamps})$ obey the required vacuum conditions. 

To determine the explicit form of the operators ${\mathbf b_i}$ we start with the observation that the states $\li\spc \mathbf{vac}\spc \ra_{\! \Hh} $ and $ \li 0 , 0\ra_{\! \Hh} $ are related by 
\be
\label{unruhvac}
\li\spc \mathbf{vac}\spc \ra_{\! \Hh} 
\is {1\over \sqrt{1+v^2}} (1+v \spc {\mathbf h}^{\dagger}_1{\mathbf h}^{\dagger}_2) \,  \li 0 , 0\ra_{\! \Hh}. 
\ee
Next, following the identical steps as in the previous subsection, we introduce the Dirac operators ${\mathbf a_1}, {\mathbf a_2}$ in terms of ${\mathbf h_1}, {\mathbf h_2}$ and ${\mathbf e_1}, {\mathbf e_2}$ exactly as given in and below eqn (\ref{acnot}).  The same arguments that lead to eqn (\ref{vaccond}) tell us now that the states (\ref{newamps}) obey
\be
\mathbf{a}_1 \li \Psi\ra= \mathbf{a}_2 \li \Psi\ra =0.
\ee
These operators act on the virtual qubits and define the creation and annihilation modes with respect to the local Minkowski vacuum.
Physical operators that act on the logical qubits will leave these combined vacuum conditions invariant. 

The state $\li \Psi\ra$ carries the required entanglement across the horizon, and gives a unique entangled state for the qubits defined by the operators ${\mathbf b_1}$ and ${\mathbf b_2}$.
This follows from the observation that ${\mathbf b_1}, {\mathbf b_2}$
are related to ${\mathbf a_1}, {\mathbf a_2}$  by the Bogolyubov transformation
\be
\label{bogol} \mathbf{b}_1 \is {1\over \sqrt{1+v^2}} \left({\mathbf a}_1 - v {\mathbf a}_2^\dagger \right), \nonumber \\[-2.5mm]\\[-2.5mm]
 \mathbf{b}_2 \is {1\over \sqrt{1+v^2}}  \left({\mathbf a}_2 + v {\mathbf a}_1^\dagger\right) \, .\nonumber
\ea
These operators represent the local Rindler operators for the infalling observer. 

We thus conclude that the firewall{\it-like} states (\ref{newamps}) in fact describe non-singular horizon geometries. The {\it real} firewall states are those in which the virtual qubits are excited and which therefore do not meet our balanced holography requirement. Note however that, even though the state (\ref{unruhvac}) describes an Unruh type vacuum with thermal entanglement, the virtual qubits themselves are not thermally excited \cite{davidlarus}. Excitations of the virtual qubits are {\it deviations} from the Unruh vacuum  state: they are 
Kruskal particles, and these do not feel any temperature.   In the microscopic theory, virtual excitations are excluded from the thermal ensemble by virtue of an energy gap.
They are still a necessary ingredient of our description, however, because  in order to represent the total entanglement across the horizon, we have to work in the 
full $2^{2N}$ dimensional Hilbert space. 

\vspace{-2mm}
\subsection{AdS/CFT embedding}

\vspace{-3mm}

We now briefly outline a proposed realization of balanced holography in AdS/CFT. Our discussion will be schematic.
Consider a strongly coupled CFT and let $| \psi\rangle$ be a typical state in the subsector ${\cal H}_{{\!}_{\rm CFT}}^E$ of 
CFT states with totel energy $M$.
 From a coarse grained perspective,  $| \psi\rangle$  looks like a thermal state
  with temperature equal to $T_H$. 
How can we probe and look behind the horizon of the dual AdS black hole geometry?

The question reveals part of the answer. Normally, one would have been inclined to think that the micro-physical units that store black hole information
reside in the region {\it behind} the horizon, not outside of it.  Indeed, historically, the real surprise of gauge/gravity duality was that the emergent space-time dual to
the CFT is filling out the complete AdS region outside the black hole. In AdS/CFT, it is evident that
black hole information is carried by hidden degrees of freedom that extend far outside the horizon --  just as in balanced holography. 
The question is how to restore the information balance between the exterior and the interior regions.

To address this question, we need the ability to examine the CFT state with localized bulk probes  \cite{paparaju}.
A natural set of probes are single trace operators with energy above some cut-off $\Lambda$,  with $T_H \ll \Lambda \ll M$. Let ${\cal H}_{{}_{\rm AUX}}$ denote the subsector spanned by tensor products of such single trace states. We assume that the state $| \psi\rangle$, when viewed as an element  of ${\cal H}_{{}_{\rm CFT}}^E \otimes {\cal H}_{{}_{\rm AUX}}$   factorizes as $|\psi\rangle_{{\! }_{\rm CFT}} \, | 0 \rangle_{{\!}_{\rm AUX}}$.
This is a reasonable restriction, since excited states in ${\cal H}_{{}_{\rm AUX}}$ are highly Boltzmann suppressed.
 
We choose the dimension of ${\cal H}_{{}_{\rm AUX}}$ to be equal to ${\cal H}_{{\!}_{\rm CFT}}^M$. To activate the auxiliary system as a probe of the state $|\psi\rangle$, we now modifiy the CFT Hamiltonian be setting it to zero inside the auxiliary subsector, $H_{{\!}_{\rm AUX}} = 0$, while leaving the interaction Hamiltonian $H_{\rm int}$ that couples to two sectors intact. The initial factorized state then becomes unstable:
the quantum information in $|\psi\ra$ will diffuse, until the system ends up in 
a maximally mixed state  \be
\li \Psi \ra =   \frac{1}{\sqrt{2^N}}\sum_{i}  {\bf U}_{i}\, \li \spc \psi \spc \ra_{{\!} _{\rm CFT }} \li \spc i \spc \ra_{{\!}_{\rm AUX}},
\ee
where the ${\bf U}_i$ depend only on the known  interaction Hamiltonian $H_{\rm int}$.
This is our proposed dual CFT representation of a balanced black hole state. 

The above recipe can be viewed as a unitary implementation of a renormalization group procedure, that integrates out all
excitations above the  cut-off $\Lambda$. In a Hamiltonian set-up, this amounts to taking the trace over the UV sector of the Hilbert space. 
The left over IR effective theory then necessarily ends up in a mixed state, with entropy equal to the entanglement entropy between the IR and UV modes
across the cut-off scale $\Lambda$. Since there are many more UV than IR modes, the entanglement is close to maximal. For most practical
applications of the renormalization group, this entanglement does not matter.
But if we want to keep track of quantum information,  we need to restore the purity of the IR state. We can do this by re-introducing the UV modes in the form of a non-dynamical purification space  ${\cal H}_{{}_{\rm AUX}}$, of the same size as the IR Hilbert space. 
The form of the purification state is determined dynamically, by turning off the UV Hamiltonian while leaving the interaction Hamiltonian across the cut-off scale $\Lambda$ in place.

\begin{figure}[t]
\begin{center}
\includegraphics[scale=.48]{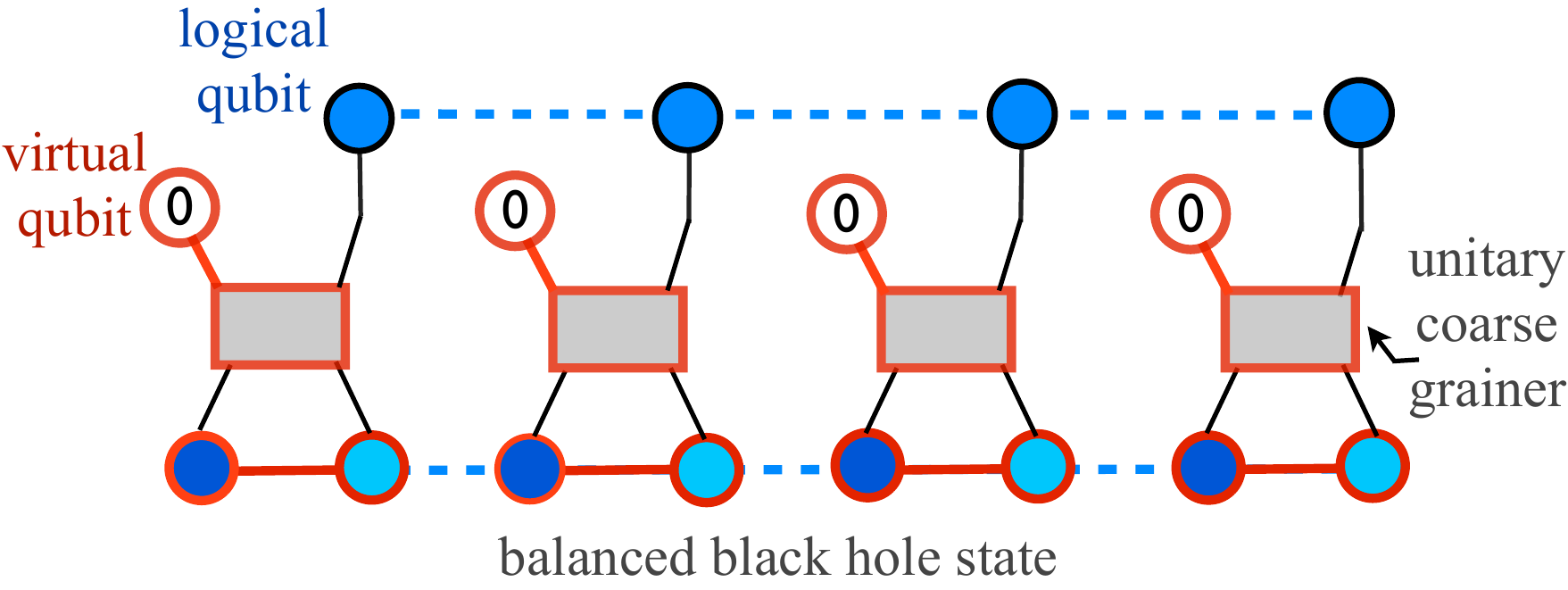}
\caption{\small The reduction from a balanced black hole state to a logical qubit state proceed via a unitary
coarse graining step, similar to that used in MERA \cite{mera}. The virtual qubits control the binary state of the coarse grainer.
The light and dark blue circles are entangled interior and exterior qubits.}
\end{center}
\vspace{-0.5cm}
\end{figure} 

The outlined RG step is very similar to the coarse graining step in MERA \cite{mera}, depicted in fig 1.
As pointed out in \cite{swingle}, the emergent dimension of AdS inherits its stability from the presence of 
coarse graining operators, with binary settings controlled by ancillary qubits analogous to our virtual qubits.

This AdS/CFT embedding perhaps clarifies the role of the virtual qubits in our earlier discussion. These virtual qubits are
supplied by the auxiliary probe system.
After mixing them in with the CFT, they provide a place holder for every possible localized probe
of the bulk geometry. Virtual excitations are generated by operators that, when translated back to before the mixing process, 
map the vacuum state $| 0\rangle_{{\! }_{\rm AUX}}$ on to some excited state.

The unitary entanglement swap that disentangles the real sector (CFT) from the virtual sector (AUX)  now has the form ${\bf U}_{{\!}_{\rm QT}} = \sum_i {\bf U}^{\dag}_i\, {\rm P}_i$, with ${\rm P}_i$ the projection on the state $|\, i \, \rangle_{{\!}_{\rm AUX}}$.  This definition is state independent: the same disentangler ${\bf U}_{{\!}_{\rm QT}}$ works for all initial states $|\psi \rangle_{{\!}_{\rm CFT}}$.
  The resulting construction~of internal operators  proceeds along similar lines as in \cite{vv}.
  
We leave further study of the CFT embedding of our proposal, and its relationship with \cite{vv},  to future work.
\vspace{-2mm}

\subsection{Conclusion}

\vspace{-2mm}

In this paper, we have formulated and utilized a new formulation of the holographic principle, which postulates that black hole
information is stored in non-local correlations between micro-physical interior and exterior  degrees of freedom. 
This proposal combines the usual assumption, that black hole information is stored inside the stretched horizon, with the point of view 
of AdS/CFT and the fuzz ball program, in which all quantum information appears to be stored in the exterior region.
It also naturally dovetails with the idea that saturation of the entanglement entropy  bound is responsible for the continuity of space.

Using a qubit representation of the proposal, we have shown that the apparent firewall obstruction can be removed,
via a universal entanglement swap operation ${\bf U}_{{\!}_{\rm QT}}$. This unitary rotation can be viewed as the
reconstruction map of the low energy effective field theory operators of the infalling observer, in terms of the microscopic variables
that encode the black hole micro-state. Relative to these observables, the horizon looks like any other smooth region of space-time.

Our discussion in this note has focussed on stationary young black holes. The original firewall argument, however, was formulated for old black holes,
that are past their Page time. It is straightforward, however, to generalize the present result to include time-dependence and old black hole states.
We refer for this discussion to the accompanying paper \cite{vv2}. In \cite{vv2} it is shown that
the pairing of the logical and virtual qubits is symbiotic: one benefits from the presence of the other. The virtual qubit protects the logical qubits from local sources of decoherence, by absorbing these into its excited state.
The logical qubits protect the virtual qubits, from having to break its local entanglement, by carrying the long range correlations  with the distant environment.
In this way black holes are capable of storing and releasing quantum information, while maintaining a locally smooth and connected horizon geometry.

\vspace{2mm}

\begin{center}
{\bf Acknowledgement}\\[1mm]
\end{center}

We thank Rafael Bousso,  Bartek Czech,  Borun Chowdhury,  Daniel Harlow, Juan Maldacena,  Samir Mathur, Kyriakos Papadodimas, Joe Polchinski,  and Douglas Stanford
 for helpful discussions.
The research of E.V. is supported by the Foundation of Fundamental Research of Matter (FOM), the European Research Council (ERC), and a Spinoza grant of the Dutch Science Organization (NWO). The work of H.V. is supported by NSF grant PHY-0756966.

\end{document}